\DocumentMetadata{}
\documentclass[sigconf,balance=false]{acmart}
\usepackage{hyperref}
\usepackage{popets}
\usepackage{threeparttable}
\usepackage{rotating}

\usepackage{siunitx}
\usepackage{graphicx}
\usepackage{booktabs}

\usepackage{fritzlistings}

\setcopyright{popets}
\copyrightyear{YYYY}

\acmYear{YYYY}
\acmVolume{YYYY}
\acmNumber{X}
\acmDOI{XXXXXXX.XXXXXXX}
\acmISBN{}
\acmConference{Proceedings on Privacy Enhancing Technologies}
\newcommand{\etal}{~\textit{et~al}.}

\begin{document}
\title{Are we collaborative yet? A Usability Perspective on Mixnet Latency for Real-Time Applications}

\author{Killian Davitt}
\affiliation{%
  \institution{King's College London}
  \city{London}
  \state{}
  \country{UK}}
\email{killian.davitt@kcl.ac.uk}

\author{Dan Ristea}
\affiliation{
  \institution{University College London}
  \city{London}
  \state{}
  \country{UK}
}
\email{dan.ristea.19@ucl.ac.uk}

\author{Steven J. Murdoch}
\affiliation{
  \institution{University College London}
  \city{London}
  \state{}
  \country{UK}
}

\renewcommand{\shortauthors}{Davitt et al.}

\begin{abstract}
Mixnet networks deliberately induce additional latency to communications to provide anonymity. 
Recent developments have allowed mixnets to reduce their latency from hours to seconds while maintaining the same level of anonymity. 
As a result, real-time communications are now possible on mixnets. 
There has been limited research on how users tolerate different levels of delay, and it is unclear what latency levels mixnet operators should choose. 
Previous studies about latency do not apply to these `mid-latency' mixnet scenarios. 
Our paper contributes the first measurement of users' tolerance to real-time applications under mixnet delay. 
We design a text-based collaborative quiz system to test user response to latency where participants complete a set of question tasks in collaboration with a simulated second user. 
Different levels of latency are added, analogous to a modern mixnet system. 
We show that average delay parameters of 1s and 4s maintain usability, a mean delay of 7s shows some difficulty and a mean delay of 10s is detrimental to user experience.

Using these delay parameters, mixnet operators can ensure that most types of real-time communication applications are usable. 
Mixnets thus can balance usability and anonymity without compromising either.

\end{abstract}


\maketitle

\section{Introduction}

The mixnet is one of the most powerful anonymity systems available to Internet users. It is one of the most longstanding privacy enhancing technologies available.
Although mixnets can provide very strong anonymity, they mostly achieve this by adding high latency to communications. 
Previous iterations of mixnets such as Mixminion~\cite{danezisMixminionDesignType2003} and Mixmaster~\cite{MixmasterProtocolVersion} worked with very high latency on the order of hours~\cite{serjantovMessageSplittingPartial2006}. 
As a result of this latency, mixnets have historically only been used for non-real-time communication, mostly emails~\cite{diazmixnetworks2019}. 
Recent developments in mixnet design provide the same anonymity guarantees with lower latency and, as such, enable the use of real-time protocols. Most modern mixnet designs are based on the classic `Stop-And-Go' design originally developed by Kesdogan\etal\cite{kesdoganStopGoMIXesProviding1998}.
Most notably, the Loopix mixnet system~\cite{piotrowskaLoopixAnonymitySystem2017} incorporates new techniques like Loop traffic and Poisson mixing that enhance anonymity guarantees. 

The lower latency required for these newer mixnets is still much greater than that offered by the Tor network~\cite{PerformanceTorMetrics}. The significant difference between average Tor latency and mixnet latencies means previous work on the usability of Tor is not applicable. Tor remains the most popular anonymity system used today~\cite{UsersTorMetrics}. Unlike Tor, mixnets are not vulnerable to a global passive adversary, which makes them desirable even with their increased latency. 

As latency level is one of the key factors affecting anonymity guarantees\cite{dasAnonymityTrilemmaStrong2018} and one of the easiest to adjust, Mixnet operators must carefully balance latency parameters with usability in their system. A system designed for non-real-time applications can use very high latency and guarantee strong anonymity. In contrast, a system designed for use with real-time applications must lower latency to the level required for usability, but anonymity guarantees may not be as strong.

So far, the usability of applications under mixnet latency has not been measured. To address this, we developed and ran a user study to evaluate users' tolerance to delay in real-time applications under latency conditions caused by modern mixnet systems. 
The types of real-time activity performed online can vary; thus, the impact of delay on users will vary by activity.

Our user study models two users collaboratively completing a contrived, but generalizable real-time typing task.
This type of task produces frequent opportunities for conflict between users and is exacerbated by higher levels of latency. 

The task was designed as a particularly difficult scenario for examination that would produce conflicts and frustration at a higher rate than any common collaboration task. 
Therefore, the study's result can be considered a relatively high bound on the difficulty induced by that delay and this upper bound applies to any type of collaborative activity.

We developed a web application to run the user study. The study's task is a simple quiz with a series of general knowledge questions that participants answer in collaboration with a simulated second user. 
Communication between the users operates under similar delay characteristics to stop-and-go mixnets. The messages in this study are delayed using samples from an exponential distribution. This ensures that this study can be used as a parallel for how users of mixnets would tolerate delay.

While collaborative applications typically use a centralized server model to transfer data, we believe this is less suitable for mixnet usage. Therefore, our system models a peer-to-peer approach. We employ the innovative Automerge software to facilitate low-latency peer-to-peer communications so that our model can be accurate but also not hindered by the added latency of the specific software used.

By testing users' responses to this typing task, we develop critical insights into users' behavior when faced with high latency. 
As a result, we provide key recommendations to modern mixnet operators on finding the appropriate latency levels to use. 
We also offer the first real context for mixnet delays and show future mixnet researchers a key metric for the usability of their systems. These results are a crucial step in making the mixnet privacy enhancing technology available for users in more diverse scenarios where it may not have been deployed before.

Our study measures four different mean latency values and does not provide a full distribution of users' responses to all possible latencies. However, the latency values selected for the study provide a useful look into the general progression of user response as latency rises.

We utilize the following research questions to inform how the study was designed to assess user's responses to this latency.
\begin{itemize}
\item{At what level of delay does collaboration slow beyond usefulness?}
\item{How does user frustration vary as delay increases?}
\item{Do users tolerate an increased delay, even after it renders the whole task slower?}
\item{Does increased delay affect task completion rates?}
\end{itemize}


\section{Background}

\subsection{Typing Speed}
\label{typing-speed}

Dhakal~\cite{dhakalObservationsTyping1362018} conducted an extensive study of typing performance characteristics on 168,000 volunteers. One key result is that participants' average typing speed is 51.56 words per minute (WPM), equivalent to 0.86 words per second. Estimating the delay per character requires the average length of words in the English language (4.7 according to Norvig~\cite{EnglishLetterFrequency}). From this data, we compute that the average user types at approximately 4 characters per second; thus, the delay between characters being typed should be 250ms. This provides a baseline for simulating the typing speed of a real user in our study.

\subsection{Collaboration Architectures}
Achieving consistency in collaborative documents is usually done using one central data authority. Typical collaborative solutions like \emph{Google Docs} are conducted over a central server, and any edits done are reconciled directly and immediately on the server. In many cases, this makes for a more simplistic and straightforward protocol than a peer-to-peer model in which the service users communicate directly. The success of services like Google Docs shows the viability of centralized services with algorithms like Operational Transformation~\cite{sunOperationalTransformationRealtime1998}; however, it also poses some additional problems for introducing anonymity. In a centralized model, common algorithms like Operational Transform require access to the data being operated on. 
Even if the participants cannot be identified based on IP address, the data in the file can potentially compromise their anonymity. Additionally, the centralized server has a unique position, allowing them access to additional network traces, which would assist in the deanonymization of participants. 

Additionally, the need for packets to traverse the central server doubles the latency in centralized settings, and given that latency is a key issue in anonymity systems, a peer-to-peer system is a sensible choice for the task. 

End-to-end encrypted centralized collaboration is possible: CryptPad~\cite{CryptPad2022} is one such solution, which uses a modified version of Operational Transform allowing clients to perform the role of the centralized server. A peer-to-peer model is nevertheless still a reasonable choice for mixnet communications.

\subsection{CRDT's}
To facilitate peer-to-peer collaboration, we employ the ``Conflict-Free Replicated Data Type'' (CRDT) data structure. A CRDT~\cite{shapiroDesigningCommutativeReplicated2007} is a data structure with only commutative operations. A CRDT can be used by collaborating users who store parallel data structure instances on their devices. Edits can be sent across a network and applied to the local CRDT, which, due to its commutative operations, will always converge to a single true state, even if the operations are applied out of order on different instances. A wide number of CRDTs have been proposed in the literature. They often store different data types and allow different operations to be performed. However, the most common CRDTs are typically used for collaborative text editing. 

\subsection{Automerge}
Automerge~\cite{kleppmannConflictFreeReplicatedJSON2017} is a project attempting to produce consistency algorithms without a central authority server. Automerge is a CRDT specifically designed for applications with poor network connectivity. 



Automerge not only guarantees the eventual consistency of collaborative documents but also performs well. We selected Automerge for the study due to its low latency and minimal overhead. This enables measuring only the effects of the artificial delay introduced without additional delay from the CRDT operations. 

While we wish to analyze a difficult collaboration scenario, we also want our model to be realistic. Employing a modern technology like Automerge ensures our results apply to the latest collaboration methods.

In addition, running real-time collaborative applications over mixnets would be impossible without modern CRDTs like Automerge and YATA~\cite{nicolaescuRealTimePeertoPeerShared2016}. 
The poor performance of older CRDTs resulted in very high latency operations or loss of critical features like deletion. 

\subsection{Appropriate Mixnet Delays}
Our study aims to measure user tolerance to the delays introduced by mixnet systems. 
To decide what level of delays should be considered, it is important to discuss the history of mixnet delays first.

Serjantov and Murdoch~\cite{serjantovMessageSplittingPartial2006} conducted some evaluations of the deployed Mixmaster and Mixminion systems in 2006 and obtained statistics on the latency experienced in the systems.
Mixmaster's median latency was 2.7 hours, and the minimum was approximately 13 minutes. 
This indicates that these early mixnet systems would be unable to support any real-time communication or collaboration systems, as the required delay is too high to make these applications usable.

In recent years, many messaging systems have been designed to prove the suitability of mixnet systems for real-time communications. 
These modern systems use different added delay techniques to enhance anonymity or use another form of artificial noise or dummy traffic. 
The net result of both techniques is increased end-to-end latency for the user's data.
Vuvuzela~\cite{vandenhooffVuvuzelaScalablePrivate2015} is a system for anonymous messaging that provides differential privacy. The system provides messaging with, on average, 37 seconds of latency.

Stadium~\cite{tyagiStadiumDistributedMetadataPrivate2017} is a real-time messaging system based on a mixnet design that provides anonymity with a delay which, in the best case, is on the order of 50 seconds. 
The Karaoke~\cite{lazarKaraokeDistributedPrivate2018} system is a similar design but with improvements to reach delays of approximately 6.8 seconds. 
Karaoke also provides a much more robust evaluation of its anonymity guarantees. At the 6.8-second latency rate, it can provide a differential privacy guarantee against deanonymisation for an estimated one year.

Atom is another anonymous system~\cite{kwonAtomHorizontallyScaling2017}, which permits microblogging. 
It offers high scalability and anonymity guarantees. However, much like older mixnet systems, it offers much higher latency communications, which can be on the order of several minutes in the best case. 
Systems like this demonstrate the usefulness of mixnets in modern systems, but underscore the more typical usage of mixnets as high latency systems.

Groove~\cite{barmanGrooveFlexibleMetadataPrivate2022} is a system based on Karaoke that provides more flexibility to users, allows them to use multiple devices, and tolerates users going offline. The trade-off for this is a higher latency at 32 seconds. 
The Yodel~\cite{lazarYodelStrongMetadata2019} system achieves voice communications in a mixnet-type system with approximately 1 second of added delay. It, however, achieves this only by working with clients who will be connected to the network a high percentage of the time and by providing weak anonymity guarantees.

All of these systems' assumptions and anonymity guarantees vary, and none can be used as a specific benchmark for the level of delay needed for robust anonymity. 
However, these examples illustrate that robust anonymity from strong adversaries cannot be achieved with a delay as low as that of mainstream anonymity systems like Tor. 
Tor provides anonymity for communications with between 200ms and 600ms of latency~\cite{PerformanceTorMetrics}. This lower level of latency is only achieved due to Tor's lack of protection against stronger adversaries like the global passive adversary. 
Higher latency levels are evidently a major factor in how these systems achieve stronger anonymity guarantees. This is the key motivation of the study, which seeks to test the usability achieved at these higher latency levels\footnote{We don't attempt to measure the full spectrum of the delays mentioned here as per reasoning in Section~\ref{delay-levels}}.

\subsection{Very Low Latency Mixnet Systems}
Piotrowska~\cite{piotrowskaStudyingAnonymityTrilemma2021} performed a comprehensive evaluation of the Nym Loopix-derived mixnet and calculated the necessary delay for the strength of anonymity required. 
The study justifies the delay parameters used by the Nym mixnet, which are low enough to rival the performance of Tor.
The study, however, only considers entropy as a measure of anonymity, which can be a useful metric but does not fully encapsulate the ability of an adversary to deanonymize users of the mixnet. Murdoch\etal~\cite{murdochQuantifyingMeasuringAnonymity2014} have previously noted the limitations of this type of measurement. 

Nym's low levels of added latency likely will not provide sufficient anonymity for the most complex adversary attacks. 
In addition, a new style of more complex deanonymisation attack is slowly developing in the literature. 
In the future, all mixnet systems may need to offer increased added latency to combat these attacks.

\subsection{Potential Improvements in Deanonymisation Attacks}
Several examples in the literature show how the more nuanced characteristics of packet flow can be leveraged to deanonymise users more efficiently. 


The MiXiM~\cite{benguiratMiXiMMixnetDesign2021} system is a method of analyzing mixnets that attempts to take account of a more complex model of communications flow, but still does not allow for a full perspective on a potential adversary's ability to deanonymize.

The MixMatch anonymity attack by Oldenburg \etal~\cite{oldenburgMixMatchFlowMatching2024} made a recent attempt to analyze mixnet traffic by analyzing the flows and patterns of traffic being sent. Most attacks prior to this considered all messages in the system to be independent. 

A study by Gaballah \etal~\cite{gaballahEffectivenessIntersectionAttacks2023} also attempted to perform de-anonymising intersection attacks on sample communications data from social media websites. The increased effectiveness of these attacks confirms experimentally the increased attack effectiveness that can be gained by an adversary when studying the patterns of communication flow rather than independent packets.

These advances are an important step in assessing adversaries' full abilities. Thus, they can guide us in selecting Mixnet parameters, making it more difficult for an adversary to deanonymize users of these systems.
The relatively low delay parameters selected for the currently deployed Nym mixnet will need to be revised for more complex attacks that may come in the future. This forms part of the motivation for examining users' response to higher delay levels in collaborative settings\footnote{Hugenroth \etal~\cite{hugenrothPoweringPrivacyEnergy} also showed the high power requirements of low latency mixnets, in particular, Nym. Increasing latency levels would make the networks more usable on mobile devices.}.

\section{Design of the Study}
\label{3:study-design}

For this study, we developed a web application that models a collaborative experience that is more tightly coupled than what users might commonly engage in. It is somewhat similar to common tasks, but produces more conflicts, and more frustration. The results of this task are therefore an approximate upper bound on the level of conflict and frustration that a user could experience in any common task.

A collaboration task was chosen as it requires coordination between parties. 
A disrupted or delayed flow of information can cause major disruption, particularly if users need to regularly adjust their actions depending on what the other users are doing. 
For example, users filling out a form collaboratively will endeavor not to duplicate work. If information flow is disrupted, users could accidentally fill out the same section twice. The unique difficulty of collaborative tasks makes the result of the study not only an approximate upper bound on frustration experienced in collaborative tasks, but potentially any non-collaborative task also.

Online collaboration takes many forms, with a variety of different possible tasks and, in particular, a variety of different levels of synchronicity. Some tasks require infrequent interaction between users. 
As this study requires a scenario where users are less tolerant of delay, we sought a task where collaborating users enter into conflict. Collaborative document editing in a word processor is a common task featuring very rapid changes; user keystrokes at their quickest can occur multiple times per second. Users can make edits very close together, increasing the likelihood of conflict. The task selected for the study should be based on collaborative document editing, but also ensure a high conflict rate. A scenario where one user edits a single paragraph and a second user edits a second different paragraph would not be appropriate as conflict is unlikely. By emphasizing conflict scenarios, we can obtain an approximate upper bound on users delay tolerance. 
A scenario with less possibility of conflict would also be more difficult to measure without a much longer-lasting study.

The constructed web application presents study participants with a series of general knowledge questions along with free text answer boxes for each question. 
Each answer box is designed to mimic a collaborative environment, and the data can be entered collaboratively. 
As the participant attempts to answer the questions, a second simulated user will also begin answering questions, thereby allowing the participant to experience how collaboration feels in the environment. 

Additional delay is be added to application data as it ``travels'' between the participant and the simulated user, thereby simulating the delays caused by using a mixnet system. 
Study participants experience 6 different sessions of this collaborative quiz, and the delay level is adjusted each time.

\subsection{Applicability of the Task}
The choice to choose a very general task means that we do not model any one particular task well. The task we use is abstract and doesn't directly model any actual scenario that users regularly engage in. The results of this study might be applied to scenarios like collaborative document editing, video games, collaborative coding, or todo lists, for example. In each of these examples, users will experience and tolerate delay in different ways. It is important that the results of this study are read with this in mind, and if the results are to be applied to a specific real world scenario it must be clearly noted what the differences are between this real-world application and our model. As our study is designed to produce a relatively intense form of delay sensitivity, most real world applications will likely fare better than what is discussed in our results.

\subsection{The Simulated Second User}
\label{sim-second}
Given that the study simulates collaboration between two users, we must provide a second user for study participants to interact with. 
However, adding an additional participant to the study would introduce a significant amount of experimental variance. Each instance of the experiment would depend on the participation of different pairs of individuals, and it would be considerably more challenging to make inferences based on the results. 
Two real participants could also have very different typing speeds, which would introduce another uncontrolled variable to the experiment.
Therefore, we create a simulated second user and use only 1 real participant. This simulated user fills out questions in a manner that accurately simulates collaborating with another real individual. 
This approach allows us to maintain consistency in the experimental conditions. 
It ensures that each participant experiences the same collaborative environment, mitigating the impact of varying user interactions on the results.

To ensure a fair simulation, the simulated user is designed to adjust their typing speed dynamically to match the study participant. Human typing speed is emulated by adding an artificial delay between the second user's keystrokes. This delay is sampled from an exponential distribution. The unpredictability of samples from the exponential distribution emulates the random variations in human typing speed. The method used to vary typing speed dynamically is discussed in Section~\ref{sim-user-description}.

The simulated user is an approximation of a real user. 
It cannot precisely model how a real user would act.  
However, this was considered an effective solution since introducing a real user also introduces other biases. 

\subsection{Collaboration Technology}
We use the Automerge~\cite{AutomergeCRDT} protocol and library to provide the collaboration model between the participant and the simulated user. 
The participant and the simulated user both use their own Automerge data structure, which stores the answers to each question as they are entered. 
After every keystroke, the participant's Automerge model is updated, and the model is shared with the simulated user. 
Similarly, whenever the simulated user makes a keystroke, it is shared with the participant. 
Automerge allows us to maintain consistency in collaboration with relatively low latency, so we can focus on the artificially added delay as the primary source of latency.

\subsection{Delay Levels}
\label{delay-levels}
The study software allows the mean delay level to be varied over different sessions, allowing users to measure their tolerance for different delays. 
The study is a within-subjects study design, meaning the same participant experiences multiple levels of delay across multiple trials of the same study scenario. 
Each participant completes six different trials. 
Each trial is randomly assigned a mean delay level from a list of options.
One of the trials is a control trial. The control trial has no second user, and thus has no delay added. This allows us to test the quality of the collaboration versus having no collaboration at all.

Stop-and-go style mixnets typically introduce message delays based on an exponential distribution, with higher mean values resulting in longer delays and providing stronger anonymity guarantees. Similarly, in our study, we implement a delay after each participant's keystroke before sending it to the simulated user. This delay is sampled from an exponential distribution, where the mean of the distribution varies in each trial.

As we have decided to model a peer-to-peer style of communication, messages are only delayed once between the sender and receiver. In a centralized server model, a delay would be added twice, once while being sent to the centralized server and again when being sent from the centralized server. To estimate the result of performing this study using a centralized server we can assume the same level of conflict and frustration with a halved delay parameter as discussed in Section~\ref{applying_multihop}.

We adopt advice from Ignat~\cite{ignatHowUserGroups2015} and avoid including delays below average typing speed given from Dhaka~\cite{dhakalObservationsTyping1362018}. Delays at that level would not provide any significant insight. 

The selected mean delays for testing are as follows: 1s, 4s, 7s, and 10s

Several factors inform this selection. 
Firstly, the lowest delay is above, but close to, the latency offered by Tor. 
Tor provides service with between 200 and 600ms~\cite{PerformanceTorMetrics} under which most types of web activity are considered usable. 
Studying these delays is not worthwhile, as results would demonstrate good usability. 
An added delay of 1 second is greater than, but still on the order of Tor latency\footnote{As discussed, the delay added to messages in the study follows an exponential distribution. The delay figures mentioned are the average result of that exponential distribution. Latency in Tor does not necessarily follow this same distribution. Thus, the latency between Tor and this system is not fully comparable}. 
Secondly, the upper limit on the mean delays is set by the length of the study task. In our study, completing all 14 questions took approximately 60 seconds. 
Any added delay exceeding a small fraction of this task's time would interfere with the task. 
For example, if the mean delay level matched the task's length, then on average, messages to and from the second user would not be transmitted by the end of the task. 
Even a mean delay of half the length would likely still cause many messages to be lost. 
Thus, using such large mean delays would not provide an accurate evaluation of the delay's effect.

Finally, the choice of mean delay was informed by a previous study of collaborative text editing. As discussed, Ignat \etal~\cite{ignatHowUserGroups2015} conducted a user study with mean delays of 4, 6, 8, and 10 seconds. From their results, these appeared to be options that showed a significant change in usability and thus would be suitable options for our study.


\subsection{Choice of Number of Mixnet Layers}
Mixnets typically consist of a number of layers L, and messages in the mixnet must pass through L mix nodes before reaching their destination. Although some common mixnet deployments include 3 layers of mixing, this is not necessarily always the ideal choice. Recent work by Ben Guirat\etal\cite{guiratMixnetOptimizationMethods2022} studies Mixnets of this style (stop-and-go mixes) with layer length from 1 to 19 nodes. Different numbers of layers are optimal depending on a number of factors, including the estimated number of compromised, adversarial nodes in the system. 

We therefore opt to consider a mixnet of 1 layer, which is most generalizable. Stop-and-go mixes delay each message through a sample from an exponential distribution. Each additional hop through a node adds another delay. An exponential distribution at mean $\lambda$ is similar to the sum of $n$ exponential distributions at mean $\frac{n}{\lambda}$. The mean of these two distributions are identical, while their variance is slightly different. This means that results from this experiment with 1 mixnet layer can be used to approximate the result from any number of layers. We discuss this further in Section~\ref{applying_multihop}. We also note that several other factors can affect latency, an increased number of mixnet layers also adds additional delay in cryptographic operations, as well as network link latency. 

Our choice introduces a limitation as the study cannot properly model the common Loopix mixnet, which uses multiple samples from an exponential distribution which is equivalent to a gamma distribution.  Our model is similar, but not identical to this.


\subsection{The Questions}
Each trial consists of 14 questions randomly sampled from a bank of pre-written questions. These questions were written to be easy to answer for most of the population.

\begin{figure}[htbp]
\begin{center}
\framebox{\includegraphics[scale=0.14]{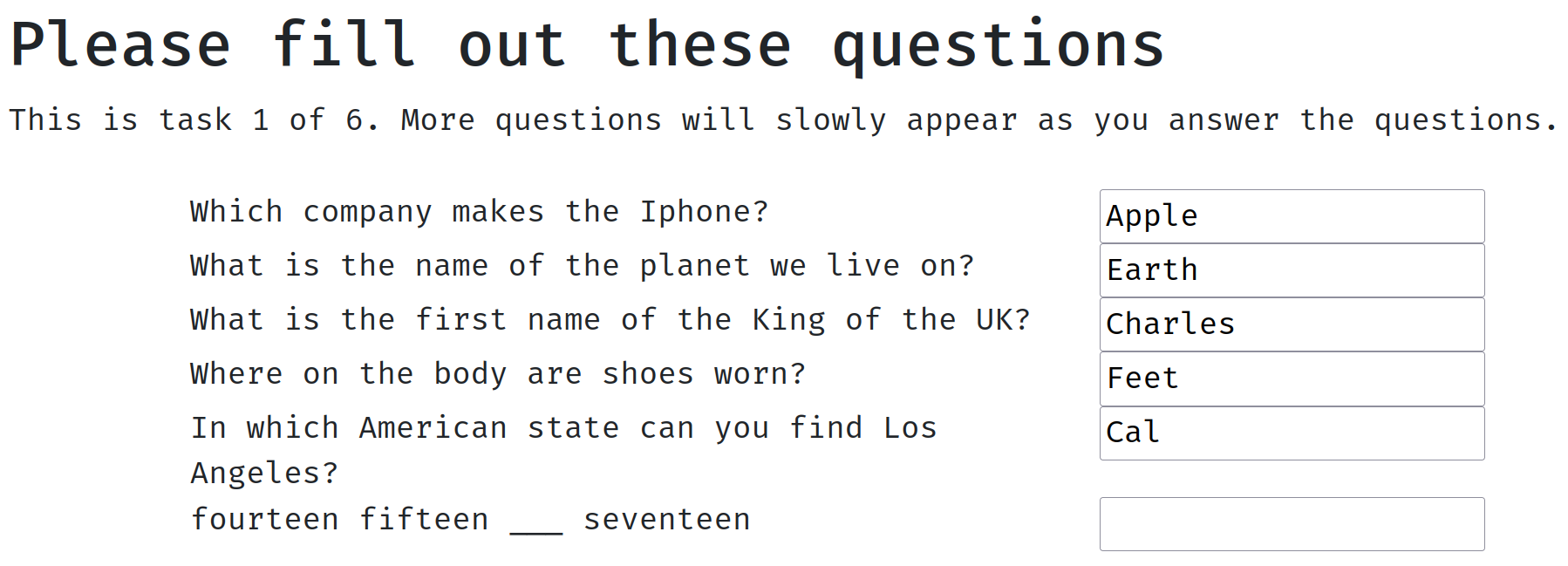}}
\caption{A screenshot from the study, about half of the questions have appeared so far in this case.}
\label{fig:study-example}
\Description[]{}
\end{center}
\end{figure}

A full list of questions can be found in the papers artifact~\cite{paperauthorsAnonymizedGithubRepository} or in Appendix~\ref{study-questions}.
A screenshot of the study application can also be seen in Figure~\ref{fig:study-example}, which demonstrates the style of questions included. 

The intention behind these questions is to provide a task for participants that is easy but requires some cognitive engagement. 
Some questions deliberately include the answer within the question, for example: ``What continent is South Africa in?''.
Other questions include prompts at the end to help participants remember the answer for example: ``Which English speaking country is nearby Australia? N\_w Z\_\_\_\_\_\_''.
The answer, ``New Zealand'', should be easily recognizable to most participants, ensuring knowledge of the answer has a negligible effect on the results.

Question selection is completely randomized, so any potentially more difficult questions have a uniform effect across the trials.

\subsection{Other Design Considerations}
We considered literature on designing effective surveys to reduce bias levels and ensure greater quality results. First, we considered the recommendations of Redmiles \etal~\cite{redmilesSummarySurveyMethodology2017a}, who presented a comprehensive summary of survey design recommendations for security and privacy research. 
By using the Prolific.com recruitment platform, it is not necessary to ask participants demographic questions directly, as this data has already been collected by Prolific and delivered to us automatically. 
This avoids biases introduced by these questions. 
We also ran pilot studies to find issues in the survey or any sections that participants did not understand.
Social desirability bias is also a key consideration for security-related studies, where users may report what they believe is the most socially desirable answer rather than what they truly would do or believe. 
This is avoided in the study by using non-self-assessed metrics as the primary result. 
Specifically, the time taken to complete the task was used as the primary metric.
Participants are not informed which level of mean delay is currently being applied, so participants cannot produce a biased result based on how disruptive they think the delay should be.

\subsection{Results Metrics} 

We focus on overall task completion time as the main metric for this study. 
Given that the simulated user is working in collaboration with the participant, the task should generally be completed more quickly than if the participant worked alone. 
Therefore, we use the time taken as a proxy for ``how helpful the simulated user was'' and ``how well the collaboration went''.
As the delay increases, the quality of collaboration is expected to diminish slowly at first. Larger delays, however, may eliminate its utility. 
If this is the case, it should become apparent in the completion time data. 
In addition, mean delay levels that cause the completion time to be significantly lower than that of the control trial (which has no simulated user) are considered to have rendered the collaboration ineffective. 

Aside from completion time, the survey also includes some self-reported measurements, which are used to give more context to the results. 
Firstly, a brief feedback question allows participants to comment on the quality of the collaboration. 
Participants also provide feedback through two Likert scale questions, which allows them to report on how frustrating the collaboration was and how much quicker they felt they could complete the task with the second user's help.
Lastly, participants are asked if they felt they had to adjust their strategy to effectively work with the second user.

We did not use the accuracy of the participant's answers as a metric. 
This would likely not provide any insight, as the results will vary depending on the user's knowledge rather than on the success of the collaboration.

\subsection{Pilot Studies}
Several pilot studies on a small number of participants were conducted to ensure the study worked as expected. 
It was not initially clear how quickly the level of user frustration would increase as the mean delay increased. Therefore, an additional purpose of the pilot studies was to obtain an insight into the levels of delay that were best to measure.

\subsubsection{Learning Effects}
Initial studies with 10 participants did not yield the expected results. While low sample sizes were not expected to show statistical significance, no relationship was found between task time and mean delay levels. This suggested a strong learning effect. Despite randomized delays, participants improved at collaborating with the second user over time, skewing the data. 

To address this, a practice round was introduced to familiarize participants with the application. The first round always included a 5ms mean delay and was otherwise unchanged. Participants were unaware that it was a practice round, and its results were excluded from the analysis.

\subsubsection{Sequential Questions}
After another pilot study yielded unexpected results, we adjusted the 2nd user's question selection method from random to sequential. Initially, random selection was chosen to mimic realistic collaboration, where users work on shared but non-overlapping tasks. However, observations showed that users naturally tackled questions sequentially. 

The new approach, where the 2nd user answered the next unanswered question in order, better reflected tightly coupled collaboration, increasing the likelihood of conflicts. This adjustment aimed to create a slightly higher collision rate than typical applications, enhancing the test's relevance. The change successfully produced more tightly coupled collaboration during testing.

This fix did not, however, provide results that demonstrated the application was working as intended. 
The qualitative comments from participants now showed that participants had developed a strategy for avoiding conflict with the second user. 
Most pilot study participants noticed that the 2nd user began at the top and sequentially completed the questions. 
Participants decided to adapt to this and began answering questions from the bottom of the list. This made conflict with the second user very unlikely until the two converged in the middle of the question list. 
This was considered a fault in the study as a realistic conflict between two users was not being measured. 
To finally fix this problem, it was decided that questions should gradually appear to the participant and the simulated user so that conflict would be much more likely between them.

One last change was made based on the pilot studies' results. Some users reported that the second user began the trial too quickly and that participants had not had a chance to read the first question before the second user began. 
We agreed that having the 2nd user begin immediately when the web page loaded was unrealistic and did not approximate real users. 
Real users require a few moments to read the first question and think of the answer. 
We added a 3-second delay to the simulated user's initiation to fix this. 
This gives the impression that the simulated user is reading the question and deciding what to write.


\subsection{Full Study Description}
The study begins with participants reading and completing the information and consent sheet. 
Participants are briefed on the nature of the study and informed how their information will be used. 
Participants are also asked for some further demographic information.

Next, there is a short question to check their understanding of the study. 
It is highlighted to them that they will be completing the survey in collaboration with a simulated user who is not an actual person. 
It is also again clarified that the purpose of the survey is not to assess users' knowledge of the questions being asked, but to see how successfully they can collaborate under increased latency. 
Participants are asked two short questions to ensure that they understand this information. 
If they do not answer correctly, they are given an additional chance to try again, and if they again do not answer correctly, they are prevented from proceeding and asked to discontinue the survey. 
This format is required by the Prolific.com platform to fairly exclude participants who have not understood the study sufficiently.

Participants then move on to the main task. 
As discussed, this task involves filling out general knowledge questions. 
While the participant completes the questions provided, the second simulated user begins to enter their responses in some of the answer sections. 

Participants experience 6 of these trials, each with a different random set of questions from the question bank and each with a different random mean delay from the six options. One of these trials is the control session and has no second user.
The first of these six sessions is the practice trial. 
This trial is identical to the others, but always occurs with a delay factor of 5ms and always happens first. 
The results of this are not used in the result calculations. 
The survey setup is unusual and unlike any task participants will likely have to complete daily. As such, participants will likely improve very rapidly in their first attempts. 
The practice round gives users time to learn the task so that the actual measured trials 2-6 provide more accurate results.

\subsubsection{Question Structure}
Initially, only 2 of the 14 questions are visible to the participant and the second user.
Whenever the answer boxes for all the currently visible questions contain some text, another two questions become visible. 
This ensures that the participant and second user will consider two questions at a time and increases the probability of collisions. 
When all 14 questions have appeared, the submit button also becomes available to the participant, and they can submit their answers when they are happy with them. 
There is no requirement for the answers to be correct before submitting. 
\vspace{5mm}
\subsubsection{Detailed Simulated User Description}
\label{sim-user-description}
The simulated user program contains a list of the questions that can be asked and the corresponding correct answers.
The program randomly selects one of the currently unanswered, visible questions to fill out and types the correct answer into the corresponding answer box. 
The exact sequence is as follows:

\begin{enumerate}

\item{Randomly select one of the visible questions that has not been answered}
\item{Type the next letter of the answer into the selected box}
\item{Sample a number from an exponential Distribution and delay for that amount}
\item{After every letter is typed, check if the current value of the answer box is a substring of the correct answer}
\item{If not, a conflict has occurred with the participant. Abandon this question and return to 1}
\item{When the answer is completed, return to 1}
\end{enumerate}


The mean of the exponential distribution in step 3 determines the typing speed of the simulated user. 
As discussed in Section~\ref{sim-second}, this mean will increase or decrease to dynamically adjust the typing speed to stay approximately in the same range as the participant user. 
The mean begins at 250 (average delay 250ms).

The dynamic typing speed is achieved by maintaining a separate typing speed comparison counter. 
This counter decreases when the second user types and increases when the study participant types. 
A large counter indicates that the participant is typing faster than the user, whereas if the counter drops to 0, the second user is typing faster than the study participant. 
The baseline typing speed matches the 250ms derived in Section~\ref{typing-speed}. 

The counter always begins at 10. 
Whenever the participant types a letter, the counter is incremented by one. 
Before the simulated user types each letter, the program checks if the count is above 1. 
If the counter is greater than one, the program types the letter and, with a probability $p=0.9$, decrements the counter by 1. 
If the count is not positive the user's typing speed is reduced by a constant factor. 
The count is then rechecked every 2 seconds, and if it remains at 0, the typing speed is reduced again.
Conversely, whenever the participant types, the program checks if the count is above 10, and if so, the simulated user's typing speed is increased by a constant factor.  
The full simulated user algorithm is shown in pseudo-code in Listing~\ref{simulated-user-code}.

This system also ensures that if the participant is not typing at all, the simulated user will also stop typing, and activity can resume once the participant begins typing again.

\begin{lstlisting}[float,language=c,caption=Simulated user algorithm,label=simulated-user-code]
while(true){
  question = selectRandomQuestion()
  if (question.isBlank()){
    while(question.NotFinished()){
      if(tokens==0){
        decreaseTypeSpeed()
        wait(2s)
      } else{
        typeCharacter()
        tokens-=1
        delay = sampleDelay()
        wait(delay)
      }
      if(question.text not in correct_answer){
        // Conflict
        // Move to a different question
        break;
      }
    }
  }
}
\end{lstlisting}

\vspace{5mm}
\subsubsection{Post Trial Survey Questions}
After the participant completes each question task, they are debriefed with questions designed to evaluate their collaboration experience. 

This debriefing survey section consists of 4 questions.
First, there is a Likert scale question. Participants are asked to react to the statement, \textit{``I found working with the second user frustrating.''} The following options are provided.

\begin{itemize}
\item{Not frustrating at all (1)}
\item{A tiny bit frustrating (2)}
\item{Very frustrating (4)}
\item{Highly frustrating, not usable (5)}
\end{itemize}

Second, another Likert scale asks: \textit{``What effect did the 2nd user have on how long it took to complete the task?''}
This aims to assess the speed at which participants thought the second user was helping them complete the task. Participants could choose a slower option if they perceived the second user slowing them down or a faster option if they felt they were not being hindered and were completing the task quicker with their help. The options are:
\begin{itemize}
\item{Much slower (-2)}
\item{Slightly slower (-1)}
\item{No change (0)}
\item{slightly quicker (1)}
\item{Much more quickly (2)}
\end{itemize} 

Next, a yes/no question asks \textit{``Did you have to adapt your actions because the 2nd user was unpredictable?''}. The thought behind this question was that users would have to adopt new strategies to deal with the increased delay, much as was described by Vaghi \etal\cite{vaghiCopingInconsistencyDue1999}.

Finally, a free text box is provided where participants can give general comments about their experience with the task. This was used in the Pilot studies to get feedback from participants.

The Likert scale questions were designed based on advice from Redmiles~\cite{redmilesSummarySurveyMethodology2017a}; only five options were presented to avoid redundant answers, and a neutral term was given, with two positive and two negative terms on either side. Two Likert scales are also presented in reverse order of positivity. The first Likert scale displays its most positive value on the left-hand side, while the second displays its most positive value on the right-hand side.

\subsection{Recruitment}
Recruitment for the study was done through the Prolific.com platform. This allowed us to obtain participants from a wide range of users who signed up for the platform to complete studies and surveys in return for payment.  Three hundred participants were recruited to participate in the survey, and the sample was balanced by sex and age, ensuring a somewhat representative sample for the study. Prolific's system provided demographic data for each participant, including age, sex, ethnicity, nationality, and employment status. Additionally, participants were asked for their general education level at the start of the study. These demographic data are used primarily to show that the study is representative of the UK population. Demographic data can be seen in Table~\ref{studydemo}.

This study received ethical approval from our institution's IRB. No major ethical issues for participants were identified. However, care was still taken to explain the study carefully and avoid any confusion for participants.

Each participant was paid \pounds1.70 for their participation. The median completion time for participants was 14 minutes and 3 seconds.
A software error caused 12 participants to face multiple trials of the same mean delay. These invalid responses were removed, leaving a total of 288 participants.

Prolific.com is generally considered to provide honest participants who provide high-quality responses to studies\cite{wangCanCrowdsourcingPlatforms2024,douglasDataQualityOnline2023,peerTurkAlternativePlatforms2017} and as such, participants attempting to trick, or `cheat' the study was not considered to be a substantial risk. Nevertheless, the data was checked for integrity. Participants with trials lasting longer than 5 minutes were manually examined to check for anomalies. 
Additionally, the time taken for all trials was analyzed, and the z-score was calculated for each one. 
No z-score exceeded 2.3. 
An excessively high z-score for any time could indicate anomalies in the data, for example, participants who left the task during one of the sessions for a break, and thus caused one of their results to be anomalously high. Alternatively, a participant who rushed the task and did not engage properly would also have an excessive z-score.

\begin{table}[htbp]
\centering
\caption{Demographics of the study participants}
\label{studydemo}

\begin{tabular}{lrlr}
\toprule
Race &  & Age &   \\
\cmidrule(lr){1-2}\cmidrule(lr){3-4}
White & 87.15\% & 18-29 & 18.75\%  \\
Asian & 4.51\% & 30-39 & 17.36\%  \\
Mixed & 1.74\% & 40-49 & 15.62\%  \\
Black & 3.12\% & 50-64 & 25.35\%  \\
Other & 0.69\% & 65+ & 22.92\%  \\
Withheld & 2.78\% &  &   \\
\hline
\hline
Education &  & Gender &   \\
\cmidrule(lr){1-2}\cmidrule(lr){3-4}

Not provided & 0.35\% & Male & 48.26\%  \\
None, or incomplete & 0.69\% & Female & 51.74\%  \\
Secondary Education & 39.93\% &  &   \\
University Degree & 38.19\% &  &   \\
Postgraduate & 20.83\% &  &   \\
\bottomrule
\end{tabular}

\end{table}

\section{Results}
\label{3:results}

\subsection{Completion Times}
As discussed, task completion time was the main metric used to test delay tolerance.
We first tested if changing the delay parameter significantly affects the time taken for tasks. This is done through a Friedman omnibus test. We run this test on the time taken for participants under mean delays of \num[round-precision = 0]{1}, \num[round-precision = 0]{4}, \num[round-precision = 0]{7}, and \num[round-precision = 0]{10} seconds. This metric does not include the control task (where no assistance was provided). The resultant test statistic is  \num[round-precision = 2]{140.38} with a p-value of \num[round-precision = 0]{3e-30}. 
This extremely low p-value shows a statistically significant effect on the time it takes participants to complete the question tasks when the delay is altered.

Next, we display the mean completion times of each delay scenario along with the control no-assistance scenario. These can be seen in Table~\ref{studycompletion}. The basic hypothesis that completion time will reduce as delay increases appears to be visible in this table. To test whether participants were tolerant of each level of delay, we performed a pair-wise statistical comparison between the task mean completion time at that delay level and the control task mean completion time. This is done using the Games-Howell posthoc test\footnote{The Games-Howell test is chosen as the different sets of completion times do not have equal variances. For more information see Games \etal~\cite{gamesPairwiseMultipleComparison1976} or Shingala \etal~\cite{shingalaComparisonPostHoc}.}.
The results are shown in Table~\ref{studyresults}. 
It can be seen that participants in the setting with a mean delay of \num{1}{\unit{\s}} complete the task significantly quicker than in the control trial, where they have no assistance (p<0.001). 
Again, it is clear in the \num{4}{\unit{\s}} scenario that participants complete the task significantly faster than in the control setting. The lower mean difference and the slightly higher p-value of 0.0025 may indicate a lessening of the effect. 
The \num{7}{\unit{\s}} scenario was not significantly different from the control trial while still maintaining a mean improvement of about \num{13} seconds in completion time. The lack of a significant difference shows that a user collaborating at \num{7}{\unit{\s}} mean delay cannot necessarily complete their task any quicker than if they were completing the task alone. Finally, the \num{10}{\unit{\s}} mean delay scenario was again not significantly different from the no assistance scenario, with an even smaller mean difference of \num{5} seconds and a p-value of 0.9. Statistical tests between the non-control tasks were not performed to reduce multiple testing correction adjustments.

Overall, the \num{10}{\unit{\s}} mean delay results in no discernible difference between completing the task without assistance and completing the task with \num{7}{\unit{\s}} likely results in no discernible difference. The lack of difference indicates that the collaboration between the participant and the simulated second user was ineffective at this delay level.

\begin{threeparttable}[htbp]
\centering
\caption{Statistical testing of participants' completion time versus the level of mean delay added. Completion times for each delay parameter are tested against the control of the no-assistance control trial. The lower delays of 1s and 4s show significant differences. The higher mean delays do not show significant differences. The Games-Howell test automatically accounts for multiple testing errors; thus, no additional correction needs to be done.}
\label{studyresults}
\begin{tabular}{@{}rrrlr@{}}
\toprule
Delay (s) & Mean Difference (s)  & t--value & p--value & Cohen's d\\
\midrule
\num[round-precision=0]{1} & \num[round-precision=2]{-23.612527778} & \num{4.134064} & \num[round-precision=5]{<0.001}***  &  \num{0.34450530156014586}\\
\num[round-precision=0]{4} &  \num[round-precision=2]{-19.675111111}   & \num{3.673402} & \num[round-precision=3]{0.0024852}*** & \num{0.3061168595626913}\\
\num[round-precision=0]{7} & \num[round-precision=2]{-13.729597222}  & \num{2.534692} & \num[round-precision=3]{0.0850467} & \num{0.21122434965490014}\\
\num[round-precision=0]{10} & \num[round-precision=2]{-5.804736556}  &  \num{0.861852} & \num{>0.9} & \num{0.07175600623979889}\\
\bottomrule
\end{tabular}
\begin{tablenotes}
	\small {
    \item[$***$]{Significant below 0.005 }
	}
 \end{tablenotes}
\end{threeparttable}

\vspace{3mm}

Cohen's d was calculated for each comparison with the results also in Table~\ref{studyresults} to show the measured effect size. 
The near 0.34 and 0.3 effect sizes for 1s and 4s indicate a small to medium effect size, confirming that the advantage of collaborating with the second user has a reasonable magnitude. 
A decreasing effect size is shown for each increase in delay. As expected, the effect size for a mean delay of \num{10}{\unit{\s}} is minimal, while the smaller delay values result in much larger effect sizes. This demonstrates that the assistance of the second user is only useful under these smaller mean delay values. We again note, however, that the effects for \num{7}{\unit{\s}} and \num{10}{\unit{\s}} were not significant, and thus, these effect sizes are merely indicative. 

\begin{table}[htbp]
\begin{center}
\caption{Mean completion times of the task. A general upward trend can be seen once delay is added, the highest level of mean delay results in a completion time almost as high as the no assistance scenario.}
\label{studycompletion}
\begin{tabular}{@{}rr@{}}
\toprule
Mean Delay (s)& Completion Time (s)\\
\midrule
\num{0} (no assist) & \num[round-precision=1]{101.56786458333333}\\
\num{1} & \num[round-precision=1]{77.95533680555556}\\ 
\num{4} & \num[round-precision=1]{81.89275347222222}\\
\num{7} & \num[round-precision=1]{87.83826736111111}\\
\num{10} & \num[round-precision=1]{95.76312802768165}\\
\bottomrule
\end{tabular}

\end{center}
\end{table}

\subsection{Likert Scale Evaluation}
Our next metric for evaluating users is the responses to the Likert scale questions as asked at the end of every session. 

\begin{figure}[htbp]
\begin{center}
\includegraphics[scale=0.5]{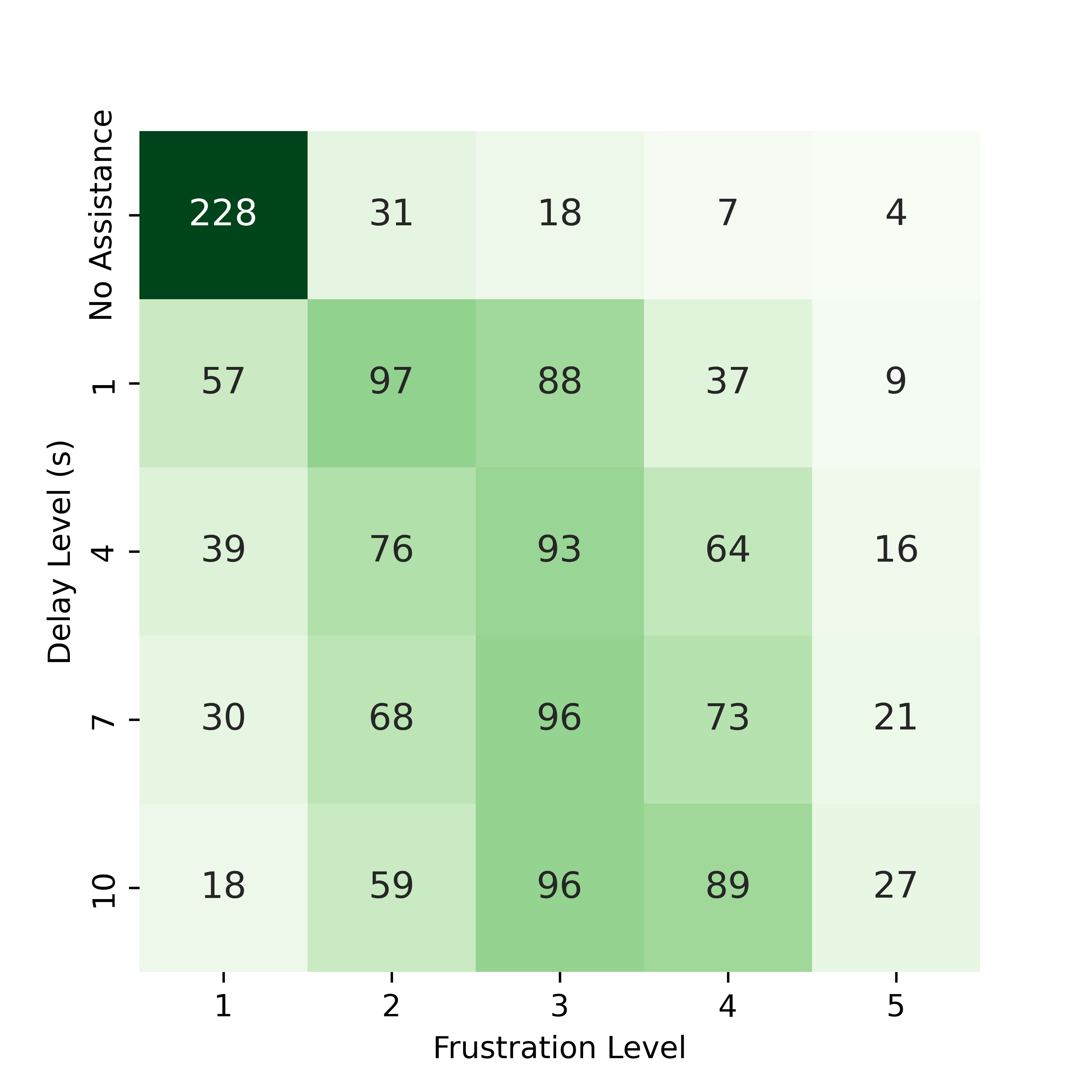}
\caption{Number of participants reporting each frustration level depending on the mean delay. An increasing level of frustration can be seen as the delay increases.}
\label{fig:frustration}
\Description[A matrix displaying the total number of users that reported each frustration level for each level of delay]{The matrix shown displays the total number of users that reported each frustration level for each level of delay. For the no assistance scenario, the number of users who reported each frustration level from 1 to 5 is 228,31,18,7,4. For the 1s scenario, the number of users who reported each frustration level from 1 to 5, respectively, is as follows: 57,97,88,37,9. For the 4s scenario, the number of users who reported each frustration level from 1 to 5, respectively, is 39,76,93,64,16. For the 7s scenario, the number of users who reported each frustration level from 1 to 5, respectively, is 30,68,96,73,21. Finally, for the 10s scenario, the number of users who reported each frustration level from 1 to 5, respectively, is as follows: 18,59,96,89,27}
\end{center}
\end{figure}

Firstly, we analyzed participants' reactions to the statement, \textit{``I found working with the second user frustrating.''} Answers could range from 1, the least frustrating, to 5, the most frustrating.
The spread of participants' answers depending on the delay level is illustrated in Figure~\ref{fig:frustration}. 
As expected, most participants chose option~1 when they did not receive assistance from a second user, as there was no user to be frustrated with. A general trend can be observed: as the level of delay increases, more participants select higher frustration levels on the questionnaire. In contrast with completion time, which remains high until the delay causes it to plateau, user frustration appears to increase constantly; thus, there may be no `sweet spot' for the delay as regards user frustration.

The results of statistical testing of these Likert questions can be seen in Tables~\ref{likert-frustration-test} and \ref{likert-speed-test}.
\begin{table}
\begin{threeparttable}[htbp]

\caption{Statistical testing of reported frustration level for participants versus the level of delay added. Reported Likert values for each delay are tested against the control of the no-assistance control. Every delay level showed significant difference from the control, with rising frustration for higher delay levels.}
\label{likert-frustration-test}
\begin{tabular}{@{}llll@{}}
\toprule
Mean Delay (s)& Mean Difference& p-value & hedges \\
\midrule
1 & -1.097 & 0*** & -1.167 \\
4 & -1.438 & 0*** & -1.479 \\
7 & -1.594 & 0*** & -1.646 \\
10 & -1.805 & 0*** & -1.909 \\
\bottomrule
\end{tabular}
\begin{tablenotes}
	\small {
    \item[$***$]{Significant below 0.005 }
	}
 \end{tablenotes}

\end{threeparttable}
\end{table}

\begin{table}
\begin{threeparttable}[htbp]
\centering
\caption{Statistical testing of perceived speed changes for participants versus the level of delay added. Reported Likert values for each delay are tested against the control of the no-assistance control. Only delays of mean 7 seconds and 10 seconds show significant differences. Higher delays resulted in participants feeling the task took longer to complete.}
\label{likert-speed-test}
\begin{tabular}{llll}
\toprule
Mean Delay (s)& Mean Difference & p-value & hedges \\
\midrule
1 & 0.024 & 0.999 & 0.02 \\
4 & 0.25 & 0.082 & 0.212 \\
7 & 0.361 & 0.003*** & 0.299 \\
10 & 0.522 & 0*** & 0.442 \\
\bottomrule
\end{tabular}
\begin{tablenotes}
	\small {
    \item[$***$]{Significant below 0.005 }
	}
 \end{tablenotes}
\end{threeparttable}
\end{table}

The second question asked of participants was, ``What effect did the 2nd user have on how long it took to complete the task?'' Answers again range from 1 to 5. This time, with 3 being a neutral value, and values on either side of that indicating a positive or negative change.

\begin{figure}[htbp]
\begin{center}
\includegraphics[scale=0.0293]{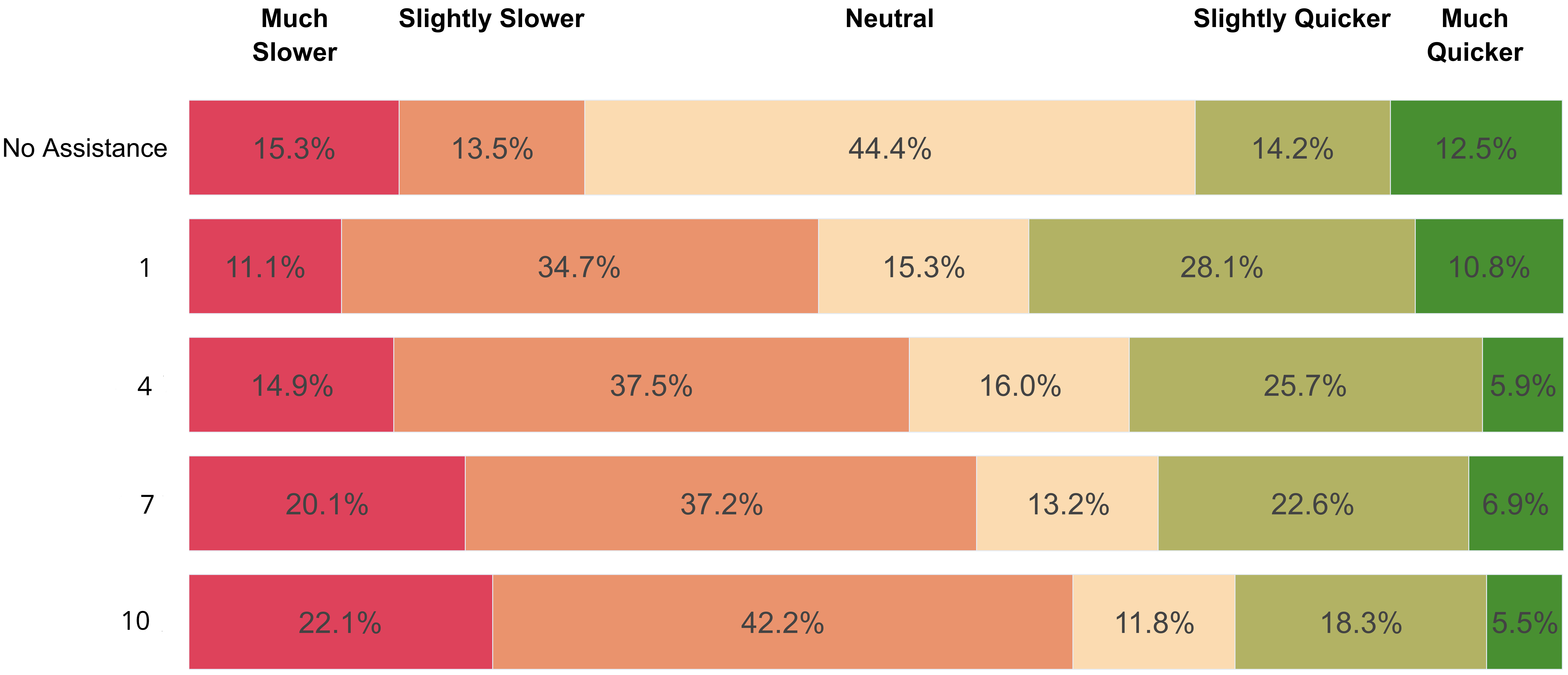}
\caption{Participant reported completion time effect from the simulated user under different mean delay levels. As the delay increased, more participants believed the second user slowed their progress.}
\label{fig:timing}
\Description[Figure shows the breakdown of what percentage of users found the experience with the second user to be slower for each level of delay.]{The matrix shown displays the percentage of users who reported whether they found the experience slower or quicker than the control for each level of delay. The options are as follows: (1) ``Much slower", (2) ``Slightly Slower", (3) ``Neutral", (4) ``Slightly Quicker", (5) ``Much Quicker". For the no assistance scenario, the percentage of users who reported each speed level as 1 to 5 is as follows: 15.3,13.5,44.4,14.2,12.5. For the 1s scenario, the percentage of users who reported each speed level from 1 to 5, respectively, are as follows: 11.1,34.7,15.3,28.1,10.8. For the 4s scenario, the percentage of users who reported each speed level from 1 to 5, respectively, are as follows: 14.9,37.5,16.0,25.7. For the 7s scenario, the percentage of users who reported each speed level from 1 to 5, respectively, are 20.1,37.2,13.2,22.6,6.9. Finally, for the 10s scenario, the percentage of users who reported each speed level from 1 to 5, respectively, are as follows: 22.1,42.2,11.8,18.3,5.5}
\end{center}
\end{figure}

The results of this question can be seen in Figure~\ref{fig:timing}. A slight trend can be seen here. Participants typically report a more negative effect from the 2nd user as the delay increases. However, the effect is minimal (The greatest change occurs once the 2nd user is introduced, which is unsurprising). 
This corroborates the previous result, namely that the advantage of working with the second user decreases as the delay rises. 
This data does not fully comport with the completion time data, however, as some participants believe that the second user is a hindrance even with low mean delays of \num{1}{\unit{\s}}. This could indicate that although users complete tasks more quickly with assistance, they sometimes perceive it as slower. 

\begin{figure}[htbp]
\begin{center}

\includegraphics[scale=0.46]{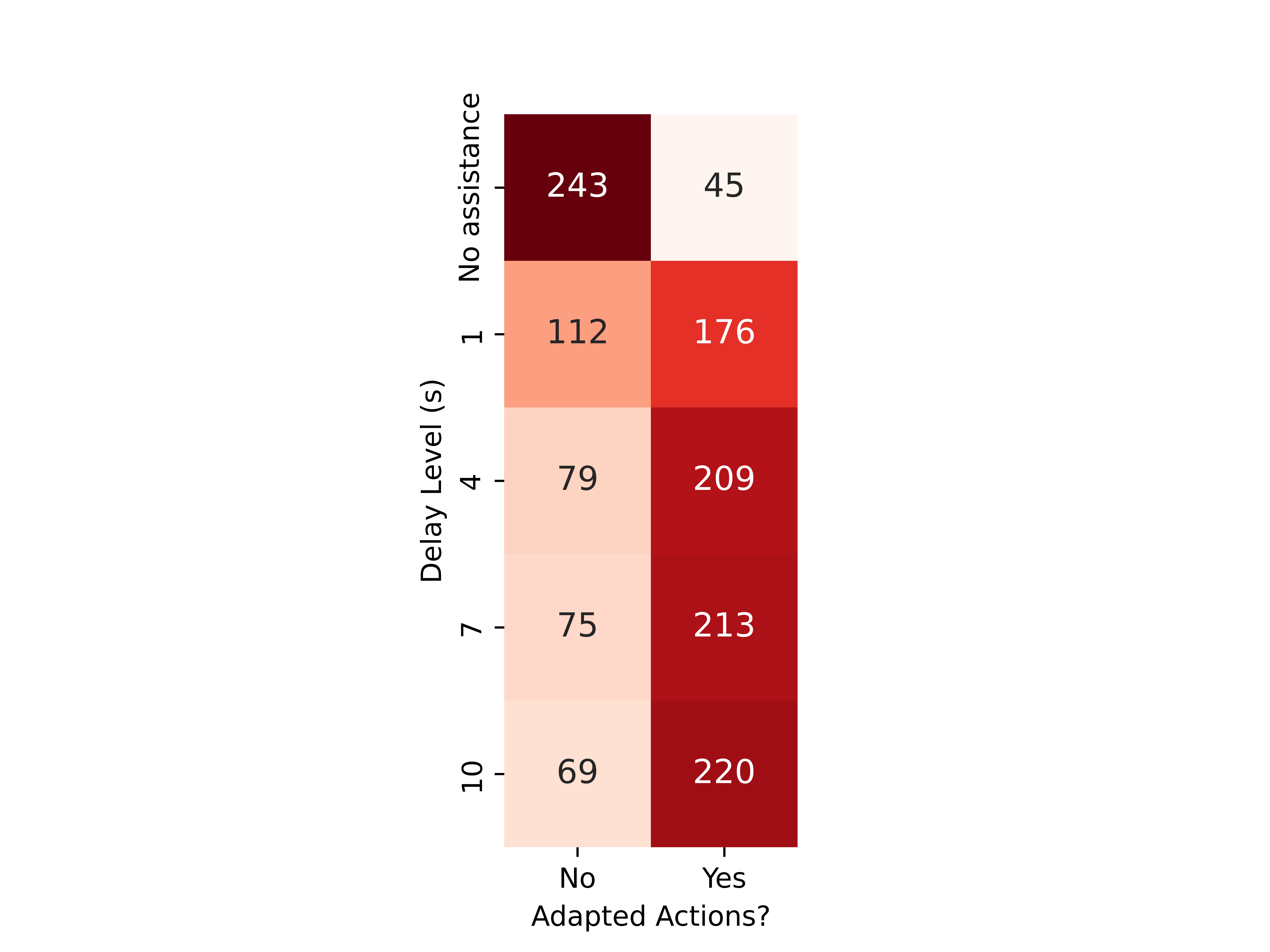}
\caption{Participant reported strategy deviation: Slightly more participants had to change strategy in some way as the delay increased.}
\label{fig:strategy}
\Description[A matrix showing how many users reported adapting their actions for each delay level]{The matrix shows the number of users who reported adapting their actions for each level of mean delay. The following are the counts of no and yes, respectively. For the no assistance scenario: 243,45. 1s 112,176. 4s 79, 209. 7s 75,213. 10s 69,220}
\end{center}

\end{figure}

Finally, participants were asked:
``Did you have to adapt your actions because the 2nd user was unpredictable?''. The results of this question can be seen in Figure~\ref{fig:strategy}. Overall, the effect is likely minor; there is a large switch from the number of users changing strategy as the 2nd user is added, but the increasing mean delay from \num{1}{\unit{\s}} to \num{10}{\unit{\s}} causes a slow increase in the number of users reporting changes of strategy. 

\section{Discussion}
\label{3:discussion}

\subsection{Excessive Delay Reduces the Speed of Collaboration}
We observed that users' task completion time increases as the mean delay increases. More specifically, it was observed that completion times at mean delay levels of \num{10}{\unit{\s}} and \num{7}{\unit{\s}} were not significantly improved upon working alone. This indicates that a mixnet that operates a real-time communication system should avoid using mean delays of \num{7}{\unit{\s}} and \num{10}{\unit{\s}}. In some scenarios, users who work together do so to increase their productivity. Two users may be able to complete the task by themselves but choose to collaborate to speed up the process. In this case, the results show that a mean delay lower than \num{7}{\unit{\s}} is essential, and even lower mean delays of perhaps even \num{1}{\unit{\s}} should be considered for maximum efficiency. However, other collaborative scenarios are formed of individuals who could not complete the task alone. Where collaboration is essential to the substance and not the task speed, a higher delay may be permissible; users will be slowed down further, but it would not be possible to complete the task otherwise.

Users' self reported collaboration speed produced a similar result. There was no significant net increase of decrease in reported completion time for mean delay levels of 1s or 4s, while mean delays of 7s and 10s showed a significant decrease in self-assessed speed. This result differs slightly from the actual recorded completion times. It is likely that while participants in the higher delay scenarios did not complete the task slower than the control task, it did feel slower due to the frustration of dealing with the second user. On the other hand, while the lower mean delay levels of 1s and 4s did complete the task quicker than the control task, self reported times are substantially mixed between slower and faster completion.

\subsection{Completion Rate is not Reduced by Increasing Delay}
Successful completion of tasks is not reduced by increasing delay, even up to mean \num{10}{\unit{\s}}. Although all other recorded variables showed great variation as the average delay varied, the overall completion rate did not change, with only a negligible number of participants abandoning the study. The study participants had a small financial incentive of \pounds 1.40 to complete the task, which should not have biased them so much as to continue the task in the face of a level of frustration that would cause them to abandon the task in normal circumstances. As a result, one can reasonably estimate that mean delays of up to \num{10}{\unit{\s}} will not cause abandonment of tasks and does not cause enough difficulty to make tasks impossible to complete. In situations where strong anonymity is of paramount importance, the mean delay could be increased to \num{10}{\unit{\s}} or possibly beyond, and users will likely still be able to complete their tasks.

\subsubsection{User Frustration Climbs Steadily with Delay}
User frustration climbs steadily as mean delay increases and may continue to increase beyond \num{10}{\unit{\s}}. Even with lower mean delays of \num{4}{\unit{\s}}, there was a marked increase in user frustration compared to \num{1}{\unit{\s}}. In contrast to earlier observations on completion times, even at lower levels of mean delay, users may become frustrated or have poorer user experiences despite still maintaining good completion times. This greater frustration reported was found to be statistically significantly different from the frustration experienced in the control setting.

\subsection{Users Often Adopt New Strategies to Assist with Difficult Circumstances}
Users often adopt new strategies for dealing with delay, which can greatly improve user delay tolerance. In the pilot studies discussed earlier, we noted that most participants approached their tasks differently after the practice round. Primarily, this took the form of participants answering questions from the bottom of the list to the top of the list. This strategy dramatically reduced the rate of collisions and conflicts with the second user. It thus negated the decreased completion time for higher delays and the frustration that would normally be associated with it. The study tasks had to be altered to avoid this type of participant adaptation as it aimed to describe a reasonably bad case for participants' conflicts. It is reasonable to expect that depending on the type of application being used, user adaption could dramatically reduce the effects of higher delay. This may occur in unexpected ways, and it is recommended to expect and test for this behavior of users.

This finding corroborates previous work and provides valuable guidance to future application developers. Allowing users to test software in realistic settings can provide insight into new user strategies for mitigating latency. This should be an important step when determining whether an application is usable with high latency, whether this be a mixnet or otherwise.

\subsection{Recommendation to Mixnet Operators}

We can recommend that when configuring mixnet systems for real-time communications, a good starting point for testing would be between \num{1}{\unit{\s}} and \num{4}{\unit{\s}}. Systems that already use an average delay lower than \num{1}{\unit{\s}} could consider opting for higher delays to bring about higher anonymity guarantees for their users. Depending on the style of communication offered by the system, completion times may not matter as much, and as a result, mean delays could be pushed much higher than \num{7}{\unit{\s}}. That being said, user satisfaction may diminish dramatically with higher delays. As is clear from the strategy of Tor Browser, user experience is critical for maintaining a large user base of diverse users. If a mean delay higher than \num{7}{\unit{\s}} is required, it should be rigorously tested for usability. In some applications, the adaptability of users to different scenarios may allow usability to be maintained in the face of higher delay and avoid this `worst-case scenario' that we have designed. This highlights a known issue of mixnet deployment: the decision to operate application-specific networks or allow general traffic. Operating an application-specific mixnet allows the operator to study their chosen application more carefully and check for user adaptation, as seen in this study's pilots.

Given that the most popular current mixnet implementation, Nym, uses much lower levels of added latency, it is hoped that future mixnet systems that aim for more robust levels of anonymity will use the recommendations provided here. As discussed, currently deployed mixnets use an adversarial model that does not fully account for the attacker's abilities. Recent work has begun to explore these more complex attacks, and it is likely that in the future, current mixnet configurations will need to be updated to defend against these attacks.

Finally, the work in this paper can inform systems that do not use deliberately added delay, such as systems that suffer from latency induced by normal network latency. Systems like Tor do not add latency directly, but the latency resulting from routing traffic across three geographically diverse nodes can reach high levels. This work informs the design of applications in systems like these. In cases where high levels of delay are non-optional, designers of collaborative applications can recommend against using their application with the system. Alternatively, they can add other latency mitigation to their user interface if it is clear that latency will exceed the levels outlined in this work.

\subsection{Applying the Result to Centralized Models and Multi-hop Mixnets}
\label{applying_multihop}
Our results can be interpreted differently to view the approximate result of a centralized model of collaboration as well as the result if mixnets with multiple hops were considered. Both of these scenarios differ in terms of the number times the messages are delayed. For mixnets of more than one hop, each additional hop would take a sample from the exponential distribution and delay the message by that amount. A centralized model of communication would then additionally double the total number of delays. We can estimate the result of these other scenarios by dividing the delay levels used by $n$, the number of delays, and viewing the same results. For example, for mean delay level 1 second we observed a mean completion time of 78 seconds. We estimate that in a 2 hop mixnet, a 78 second completion time would be achieved when applying 500ms of mean delay. This method is not exact as a single sampling from an exponential distribution at mean $\lambda$ is not exactly equivalent to sampling $n$ times at mean $\frac{n}{\lambda}$. The resulting distributions have identical mean, with slightly differing shapes, making this a highly effective, but not perfect estimate.

\section{Limitations}

\subsection{Learning Effects}
As Vaghi~\cite{vaghiCopingInconsistencyDue1999} describes, users, when faced with excessive delay in collaborative applications, can develop new strategies to avoid problems. As users complete five separate trials in the study, they may learn similar methods of avoiding conflict. We compensated for this delay by randomizing the order in which the delays occurred. If users do indeed learn, each delay level will suffer from this bias in equal measure, thereby negating its effect on the results.

\subsection{Collaboration Scenario Does not Fully Mimic Reality}
The study's designed scenario attempts to be a good approximation of actual user collaboration, but it is not identical. The task of entering questions as the study is designed does not exactly duplicate any known real-world collaborative application. While we argue that it provides a good approximation of a worst-case collaborative session, there may be details that are left out.

\subsection{Limited Abilities of Simulated User}
The simulated second user that the study participants interact with is also not a replica of a real user. As Vaghi~\cite{vaghiCopingInconsistencyDue1999} reminds us, users often develop unique strategies for avoiding conflict in collaborative work. These strategies can involve both users working together to avoid conflict. The simulated user does not allow this and cannot learn or work with the participant to avoid conflict. 
The study's results show that user strategies such as this could slightly raise the potential maximum tolerable induced delay.

\subsection{Limit on Upper Delay}
Because of the short nature of the trials used, there is an upper bound on the level of delay that can be measured. In the worst case, as delays reach the time it takes for a participant to complete the trial, it will mean that participants complete the trial fully before any interaction is seen from the simulated user. This would not provide any useful data, as the users will not interact. In other cases, even when the delay is not the full length of the trial, the time the participant spends not being influenced by interactions with the second user can be significant and is not amortized properly in these short trials.

\subsection{Latent Network Latency is Not Considered}
Real-world traffic on a mixnet not only includes the deliberately induced mixnet latency, traffic also suffers from network latency like any other traffic. This delay is not included in our study as network latency is unpredictable and will vary substantially from person to person. Generally, that network latency is expected to be considerably lower than the latency induced by the mixnet, however, real network latency can be unpredictable and can depend on a number of other factors. Readers should consider how the specific circumstances of their deployment (e.g. geographic location of users, quality of their connections) could effect the network latency, and thus the applicability of the results of this study. As an example, the Tor network does not induce additional delay by design, the observed 200-600 millisecond latency in the network\cite{PerformanceTorMetrics} can be used as an estimate for both network delay and message processing delay for a 3-hop mixnet; for a single hop, this could lie between 100ms and 300ms. While this is significantly lower than the delay levels being induced in this study, this is the mean data from Tor and specific instances and users can encounter very different circumstances. We also additionally note that some additional delay will be introduced due to cryptographic operations and that has not been modelled by this study.

\subsection{Result is Applicable Generally, not Specifically}
We attempted to model a difficult scenario for users of anonymous communications systems. The collaborative setting sought the situation where delays would most affect participants. We also attempted to make participants conflict with the simulated users as much as possible. This is valid when determining a loose lower bound on user experience in that delay, but it does not provide results that are specific to any one scenario. It is conceivable that users will be vastly more tolerant of delay in other situations such as general web browsing. As such, the user study only shows an approximate lower bound on user tolerance.

\section{Related Work}
\label{3:related-work}

The user study we designed on the usability of anonymous applications is novel. We consulted related studies in the literature, which are outlined here with notes on their commonalities and points of interest, which assisted in the design of our study.

\subsection{Studies on Text Editing}
Ignat \etal~\cite{ignatHowUserGroups2015} presented an important example of a user study focused on collaborative applications. The study aimed to evaluate how the quality of users' prose is affected by delays introduced in a collaborative text editing environment. The task chosen by the authors is influenced by conclusions from Olsen and Olsen~\cite{olsonDistanceMatters2000}, who described how closeness in two collaborators' tasks causes more problems in collaboration. Tasks with close interaction between participants are said to have high coupling. In collaborative note-taking, participants edit text that is reasonably near to each other, but not directly beside, thereby making it a medium coupling task as regards conflict. Ultimately, the authors find a strong correlation between user errors and the level of latency introduced.
The authors also note that any delay introduced that is less than the participants' average typing speed is unlikely to affect satisfaction levels and can be discounted.

Kopsell experimented with tolerance for delay in anonymity systems~\cite{kopsellLowLatencyAnonymous2006}. The experiment was performed on the JAP anonymity network, which was a mixnet-style anonymity network. At the time of experimentation, most of the traffic for this network traveled through servers operated by the authors. The authors could periodically add additional artificial delay to their servers and monitor the reaction from users. The results, over many months, clearly show a linear relationship between the number of seconds of additional delay and the decrease in the number of users of the service. 


\subsection{Impact of Delay in Other Scenarios}

Bai \etal~\cite{baiUnderstandingLeveragingImpact2018} studied users' latency tolerance in web search. The study finds that less than 500ms of delay is not reliably detectable by users, and greater than 1000ms of delay is reliably detectable by users. 

Arapakis \etal~\cite{arapakisImpactResponseLatency2021} completed a similar study about web search. 
The authors conducted an experiment to test participants' responses under different levels of delay. 
Participants were given one of a set of complex research tasks that involved investigating a particular topic and finding a list of items in that area. 
The tasks were designed to require a number of different search queries. Different levels of delay were added to the search engine for each task. 
The authors found that mobile users are more tolerant than desktop users to latency, but that, nevertheless, after approximately 7-10 seconds of delay, users show signs of negative feelings and an overall less satisfactory experience.




Nah \etal~\cite{nahStudyTolerableWaiting2003} discuss the time users are willing to wait for a web page to load. 
The study involved approximately 40 participants who took part in various web page loading simulations. 
The study evaluates users' tolerance for loading speed using an exact metric: how long a user waits until they give up and close the web page. 
The study shows that there is an obvious cutoff point where it is clear a user's tolerance has reached its limit. 

Galletta \etal~\cite{gallettaWebSiteDelays2004} also conducted a study on tolerance of website delays. Generally, the study confirms previous findings that there is a significant effect on users from increased delay, but also the study showed that a wide variety of factors can affect this mechanism, including the type of website task as well as users' familiarity with that task.

Van Damme \etal~\cite{vandammeImpactLatencyQoE2024} studied the impact of latency up to 500ms on a virtual reality collaborative task. Participants reported more difficulty with higher levels of latency. Participants were faced with different types of latency, primarily using differences in the burst levels of the latency. Participants had difficulty distinguishing between high and low levels of constant delay with high bursts of latency.

\subsection{Effect of Delay is Task Dependent}
Beigbeder~\cite{beigbederEffectsLossLatency2004} showed how increased latency affected players of the Unreal Tournament First Person Shooter game. 
They concluded that different interactions with the game were susceptible to increased latency in different amounts. 
Moving around the game world, for example, was not significantly affected by increased delay or significant packet loss. 
However, aiming and shooting were dramatically affected, with even 100ms delays decreasing users' accuracy by 50\%.

\subsection{Users Come up with Strategies}
Vaghi \etal~\cite{vaghiCopingInconsistencyDue1999} experiment with user tolerance of delay in a simple networked game. 
Two study participants controlled the simple football-style ball game. 
Different levels of delay were added for each pair of participants, and the result was noted. 
The authors found the players were employing a series of interesting coping strategies or techniques. Some players attempted to slow the ball down to reduce the delay's relative effect on its whereabouts. Some players used long-term predictions of the ball's trajectory. Users finding different strategies to avoid the consequences of delay are likely to generalize to other forms of collaboration. Similar adaptation may occur in other forms of collaboration.

\subsection{Application Design Choices Mitigate Delay}
Various techniques have been used in online collaborative applications to alleviate the effects of latency. Often, these are psychological tricks to make users more tolerant of latency issues. 
Savery \etal~\cite{saveryItTimeConfronting2011} highlights a software prediction technique that alleviates latency caused frustration in the online game \textit{World of Warcraft}.

Gutwin~\cite{gutwinRevealingDelayCollaborative2004} proposed using `Traces' to enhance the usability of collaborative applications. The previous actions of collaborators can be displayed to users in the hope that users can more effectively predict where the user will act next, thereby neutralizing some of the negative latency effects.
One example given by Gutwin is a user's cursor.
A trace, or shadow, of where the user's cursor has previously been is displayed, allowing them to predict the other user's intentions to some extent and predict where they will move their cursor next.
The techniques employed show significant results; users were better at predicting where a user's cursor would end up if the decorator was added, telling them the delay at each update point.


Yeh \etal~\cite{yehEffectsUpdateInterval2024} studied two different methods of improving user experience with collaborative document editing by delaying the display of users' edits to other users. Several interventions are studied, for example, delaying characters from being displayed until they form words or delaying sentences from being displayed until they are completed. 

Khalid \etal~\cite{khalidInvestigatingEffectNetwork2023} completed a study investigating mitigations for delay in a collaborative virtual environment. Several visual tweaks are made as a delay ranging from 50ms to 200ms is added.


\section{Conclusion}
In this paper, we outlined the problem of delay in the usability of collaborative actions. We described how the study was designed to create a scenario that maximized the effect of delay on users, which was a collaborative quiz. We discussed how the study was deployed and run. We organized and displayed the study results and made conclusions based on this data. It was shown that delays up to \num{7}{\unit{\s}} could be reasonable for mixnet operations, and results greater than \num{10}{\unit{\s}} will likely result in collaboration breakdown. Finally, we discussed the implications for mixnet operators based on these results. Overall, this paper has produced the first modern work showing what kind of delay might be acceptable for users of mixnets for real-time applications. It can potentially feed a discussion for future work, bringing mixnets from their previous life as asynchronous communication mechanisms to real-time modern applications. These advances can allow mixnet privacy enhancing technologies to be deployed more widely in the future and provide anonymity guarantees to users in more scenarios.

\begin{acks}
Killian Davitt and Steven J. Murdoch are funded by the Royal Society (RGF\textbackslash{}EA\textbackslash{}80191 and UF160505).
Dan Ristea is funded by UK EPSRC grant EP/S022503/1 supporting the CDT in Cybersecurity at UCL.

\end{acks}

\bibliographystyle{ACM-Reference-Format}
\bibliography{references}

\appendix
\section{Questions used in the user study}
\label{study-questions}
What is the capital of France,Paris\\
one two \_\_ four  five,Three\\
fourteen fifteen \_\_\_ seventeen,Sixteen\\
The United States of \_\_,America\\
What is seven plus seven,fourteen\\
What is twenty plus twenty,forty\\
What is twenty minus five,fifteen\\
Who is the UK Prime Minister,Rishi Sunak\\
Who is the president of the United States?,Joe Biden\\
What is eight minus four:,four\\
What language is spoken in England?,English\\
What language is spoken in Germany?,German\\
What language is spoken in Japan?,Japanese\\
What is the capital of Italy?,Rome\\
What is the capital of the UK?,London\\
What is the capital of China?,Beijing\\
What is the capital of Japan?,Tokyo\\
What is a three sided shape called?,Triangle\\
Which country is famous for the "Great Wall",China\\
Which month comes after July?,August\\
What colour is a banana,Yellow\\
What colour is the sky?,Blue\\
What is the name of the planet we live on?,Earth\\
What shape has four sides?,Square\\
What colour is snow?,White\\
What colour is grass?,Green\\
What is the opposite of day?,Night\\
Which orange fruit is commonly made into juice?,Orange \\
What is London's biggest airport,Heathrow\\
What bag is used to make tea?,Tea Bag\\
Which company makes computer the software 'Windows' 'Office' and 'Outlook',Microsoft\\
Which city in Germany was divided by a wall until 1989?,Berlin\\
Which city is the headquarters of most European Union organisations?,Brussels\\
Which orange vegetable is commonly eaten in the UK?,Carrot\\
Which alcoholic beverage is made from grapes?,Wine\\
Which country is north of the USA?,Canada\\
What continent is South Africa in?,Africa\\
In which country would you find the pyramids?,Egypt\\
What is the biggest country in the UK?,England\\
What is the north-most country in the UK?,Scotland\\
Which country is the Eiffel Tower located in?,France\\
In which city can you find the Statue of Liberty?,New York City\\
In which American state can you find Los Angeles?,California\\
Which animal is Wool produced from?,Sheep\\
Which currency is used in the UK?,Pound Sterling\\
Which company makes the Iphone?,Apple\\
Name the large British city: Ma\_\_\_\_\_\_\_\_,Manchester\\
In San Francisco you can find the: G\_\_\_\_\_ G\_\_\_ Bridge,Golden Gate Bridge\\
The capital of Scotland is: Ed\_\_\_\_\_\_,Edinburgh \\
Complete the sequence: Up Down Left \_\_\_\_\_,Right\\
Complete the sequence: North South East \_\_\_\_,West\\
Which organ of the body pumps blood?,heart\\
Where on the body are shoes worn?,Feet\\
Which organ is responsible for smell,Nose\\
What is the first name of the King of the UK?,Charles\\
Which Marvel superhero has the powers of a Spider?,Spiderman\\
Which country is referred to as "down under": Au\_\_\_\_\_\_\_,Australia\\
Which British author wrote Hamlet? Wi\_\_\_\_\_ Sha\_\_\_\_\_\_\_\_,William Shakespeare\\
Which English speaking country is nearby Australia? N\_w Z\_\_\_\_\_\_,New Zealand\\
Which white coloured drink is obtained from Cows?,Milk\\
In which Palace does the British King live in London?,Buckingham Palace\\
Which famous clock is attached to the houses of parliament in London?,Big Ben\\
There are two main types of smartphone: Iphone and A\_\_\_\_d,Android\\
On which social network are 'Tweets' found?,Twitter\\
Which social network was founded by Mark Zuckerburg?,Facebook\\
Which social network is commonly referred to as 'Insta'?,Instagram\\
Which video streaming service has a red 'N' as its logo?,Netflix\\
Which large online retailer shares it's name with a large river in Brazil? A\_\_\_\_\_,Amazon\\
Which liquid falls from the sky as rain?,Water\\
Which ocean separates Europe and America? Atl\_\_\_\_\_,Atlantic\\
Which large ship famously sank in 1912 after hitting an iceberg? Ti\_\_\_\_\_,Titanic\\
What language is spoken in Portugal?,Portuguese\\
Which Ocean is on the west coast of the USA?,Pacific\\
What do caterpillars turn into?,Butterfly\\
What phenomenon makes things fall to the ground?,Gravity\\
What white fluffy thing appears in the sky?,Clouds\\
Who was the first man on the moon?,Neil Armstrong\\
Which very popular messaging app has a green logo?,WhatsApp\\
Which superhero has the appearance of a bat?,Batman\\
What item worn on the hand/arm tells the time?,Watch\\
Which item do some people wear on their eyes to improve their eyesight,Glasses\\
Which commonly eaten sweet food is typicaly brown and comes in bar form?,Chocolate\\
What language is most spoken in the USA?,English\\
Which language is spoken in Spain?,Spanish\\

\end{document}